\documentclass[aps,prl,twocolumn,superscriptaddress,floatfix]{revtex4}%
\usepackage{graphicx}
\usepackage{dcolumn}
\usepackage{bm}
\usepackage{hyperref}
\usepackage{amssymb}
\usepackage{amsmath}
\usepackage{amsfonts}
\usepackage{amssymb}
\usepackage{color}%
\begin{document}
\title{Preparation of a quantum degenerate mixture of $^{23}$Na$^{40}$K molecules and $^{40}$K atoms}
\author{Jin Cao}
\thanks{These authors contributed equally to this work.}
\affiliation{Hefei National Research Center for Physical Sciences at the Microscale and School of Physical Sciences, University of Science and Technology of China, Hefei 230026, China}
\affiliation{Shanghai Research Center for Quantum Science and CAS Center for Excellence in Quantum Information and Quantum Physics, University of Science and Technology of China, Shanghai 201315, China}
\author{Huan Yang}
\thanks{These authors contributed equally to this work.}
\affiliation{Hefei National Research Center for Physical Sciences at the Microscale and School of Physical Sciences, University of Science and Technology of China, Hefei 230026, China}
\affiliation{Shanghai Research Center for Quantum Science and CAS Center for Excellence in Quantum Information and Quantum Physics, University of Science and Technology of China, Shanghai 201315, China}
\author{Zhen Su}
\affiliation{Hefei National Research Center for Physical Sciences at the Microscale and School of Physical Sciences, University of Science and Technology of China, Hefei 230026, China}
\affiliation{Shanghai Research Center for Quantum Science and CAS Center for Excellence in Quantum Information and Quantum Physics, University of Science and Technology of China, Shanghai 201315, China}
\author{Xin-Yao Wang}
\affiliation{Hefei National Research Center for Physical Sciences at the Microscale and School of Physical Sciences, University of Science and Technology of China, Hefei 230026, China}
\affiliation{Shanghai Research Center for Quantum Science and CAS Center for Excellence in Quantum Information and Quantum Physics, University of Science and Technology of China, Shanghai 201315, China}
\author{Jun Rui}
\affiliation{Hefei National Research Center for Physical Sciences at the Microscale and School of Physical Sciences, University of Science and Technology of China, Hefei 230026, China}
\affiliation{Shanghai Research Center for Quantum Science and CAS Center for Excellence in Quantum Information and Quantum Physics, University of Science and Technology of China, Shanghai 201315, China}
\affiliation{Hefei National Laboratory, University of Science and Technology of China, Hefei 230088, China}
\author{Bo Zhao}
\affiliation{Hefei National Research Center for Physical Sciences at the Microscale and School of Physical Sciences, University of Science and Technology of China, Hefei 230026, China}
\affiliation{Shanghai Research Center for Quantum Science and CAS Center for Excellence in Quantum Information and Quantum Physics, University of Science and Technology of China, Shanghai 201315, China}
\affiliation{Hefei National Laboratory, University of Science and Technology of China, Hefei 230088, China}
\author{Jian-Wei Pan}
\affiliation{Hefei National Research Center for Physical Sciences at the Microscale and School of Physical Sciences, University of Science and Technology of China, Hefei 230026, China}
\affiliation{Shanghai Research Center for Quantum Science and CAS Center for Excellence in Quantum Information and Quantum Physics, University of Science and Technology of China, Shanghai 201315, China}
\affiliation{Hefei National Laboratory, University of Science and Technology of China, Hefei 230088, China}

\begin{abstract}{We report on the preparation of a quantum degenerate mixture of $^{23}$Na$^{40}$K molecules and $^{40}$K atoms. A deeply degenerate atomic mixture of $^{23}$Na and $^{40}$K atoms with a large number ratio ($N_F/N_B\approx 6$) is prepared by the mode-matching loading of atoms from a cloverleaf-type magnetic trap into a large-volume horizontal optical dipole trap and evaporative cooling in a large-volume three-beam optical dipole trap. About $3.0\times10^4$ $^{23}$Na$^{40}$K ground-state molecules are created through magneto-association followed by stimulated adiabatic Raman passage. The 2D density distribution of the molecules is fit to the Fermi-Dirac distribution with $T/T_F\approx0.4-0.5$. In the atom-molecule mixture, the elastic collisions provide a thermalization mechanism for the molecules. In a few tens of milliseconds which are larger than the typical thermalization time, the degeneracy of the molecules is maintained, which may be due to the Pauli exclusion principle. The quantum degenerate mixture of  $^{23}$Na$^{40}$K molecules and $^{40}$K atoms can be used to study strongly interacting atom-molecule mixtures and to prepare ultracold triatomic molecular gases. }
\end{abstract}
\maketitle

Ultracold molecules have attracted great interest due to their potential applications in the study of chemical reactions at nearly absolute zero temperatures, quantum simulation of exotic models in condense matter physics, and precision measurements \cite{Krems2008,Quemener2012,Carr2009}. Recently, magnetic tunable atom-molecule Feshbach resonances have been observed in elastic and inelastic collisions between $^{23}$Na$^{40}$K molecules and $^{40}$K atoms \cite{Yang2019,Yang2022,Wang2021,Su2022}, and reactive collisions between $^{23}$Na$^{6}$Li molecules and $^{23}$Na atoms \cite{Son2022}. The analysis in Ref.\cite{Wang2021} suggests that these resonances are  a common feature of atom-molecule systems and thus provide a powerful tool to control atom-molecule interactions \cite{chin2010}. The ultracold atom-molecule mixtures with tunable interactions open up many new research opportunities, such as the adiabatic magneto-association of triatomic molecules \cite{Koehler2006,Hermsmeier2021}, the Efimov resonances involving molecules \cite{Kraemer2006}, the study of the angulon problem \cite{Schmidt2015,Schmidt2016} and the superfluid involving atom-molecule pairs. However, many of these applications require the preparation of a quantum degenerate atom-molecule mixture, which remains elusive.

Although the techniques of preparing quantum degenerate gas of atoms are well established, the creation of quantum degenerate gas of molecules is very difficult. Recently, various diatomic molecules have been created via ultracold association \cite{Ni2008,Molony2014,Takekoshi2014,Park2015,Guo2016,Rvachov2017,seesselberg2018,LiuL2019,Voges2020} and direct laser cooling \cite{Barry2014,Cheuk2018,Caldwell2019,Ding2020}. However, only two groups can create quantum degenerate gas of molecules \cite{DeMarco2019,Giacomo2020,Duda2021,Schindewolf2022}. In the experiment at the JILA group \cite{DeMarco2019}, a deeply degenerate Bose-Fermi mixture of $^{87}$Rb and $^{40}$K atoms with a large number imbalance is prepared to form the $^{87}$Rb$^{40}$K  molecules in a conventional optical dipole trap. For a large number ratio of $N_F/N_B\approx 7$ with $N_B$ ($N_F$) the number of bosonic (fermionic) atoms, the spatial mismatch of density between Bose-Einstein condensates (BEC) and degenerate Fermi gases (DFG) is mitigated, and thus a high efficiency of association of molecules can be achieved.  In the experiment at the MPQ group \cite{Duda2021}, a deeply degenerate Bose-Fermi mixture of $^{23}$Na and $^{40}$K atoms with a small number imbalance ($N_F/N_B\approx 3$) is prepared. To form molecules with a high efficiency, a species-dependent dipole trap is used to achieve density-matching between BEC and DFG.

To prepare a quantum degenerate mixture of $^{23}$Na$^{40}$K molecules and $^{40}$K atoms, the formation of molecules in a deeply degenerate mixture of $^{23}$Na and $^{40}$K atoms with a large number imbalance is desired, since it does not require the complicated species-dependent dipole trap, and the Feshbach molecules can be quickly transferred to the ground state once formed. However, it is challenging to prepare a deeply degenerate mixture of $^{23}$Na and $^{40}$K atoms with a large number imbalance, because the three-body loss rate between $^{23}$Na and $^{40}$K is about 10 times larger than that between $^{87}$Rb and $^{40}$K \cite{Duda2021}. Another difficulty in the study of atom-molecule mixtures is that the molecules usually suffer from losses due to collisions with the atoms. Different mechanisms may contribute to the losses, such as photoexcitation by trap lasers, inelastic collisions, and sticky collisions \cite{Christianen2019,Liu2020,Gregory2020,Nichols2022,Mayle2012,Matthew2021}.  In the mixtures, the elastic collisions must be faster than the inelastic collisions, so that the collisional interactions can be studied before notable losses take place.

In this Letter, we report on the preparation of a quantum degenerate mixture of $^{23}$Na$^{40}$K molecules and $^{40}$K atoms. We first prepare a deeply degenerate mixture of $^{23}$Na and $^{40}$K atoms with a large number imbalance ($N_F/N_B\approx 6$). This is achieved by the mode-matching loading of atoms from a cloverleaf-type magnetic trap into a large-volume horizontal optical dipole trap and performing evaporative cooling in a large-volume three-beam optical dipole trap.  We typically create approximately $3.9\times10^5$ $^{40}$K atoms with $T/T_F\approx0.2-0.25$ at a temperature of about 120 nK coexisting with approximately $6.2\times10^4$ $^{23}$Na atoms with a BEC fraction of about 80\%. About $4.3\times10^4$ $^{23}$Na$^{40}$K Feshbach molecules can be formed by adiabatic magneto-association. For a typical STIRAP efficiency of 70\%, about $3.0\times10^4$ $^{23}$Na$^{40}$K ground-state molecules can be prepared. The 2D column density distribution of the molecules is fit to the Fermi-Dirac distribution with $T/T_F\approx0.4-0.5$.  By selectively removing $^{23}$Na atoms, we obtain a quantum degenerate mixture of $^{23}$Na$^{40}$K molecules and $^{40}$K atoms. The elastic collisions between them provide the thermalization mechanism for the molecules. After a few tens of milliseconds which is larger than the typical thermalization time, the momentum distribution of the molecules can still be fit to the Fermi-Dirac distribution with $T/T_F\approx0.5-0.6$. This indicates that the low entropy of the molecular gas can be maintained in the mixture, which may be caused by the suppression of collisions due to Pauli exclusion principle. The preparation of a quantum degenerate atom-molecule mixture paves the way toward studying strongly interacting atom-molecule mixtures and preparing ultracold triatomic molecular gases.

\begin{figure}[ptb]
\centering
\includegraphics[width=8cm]{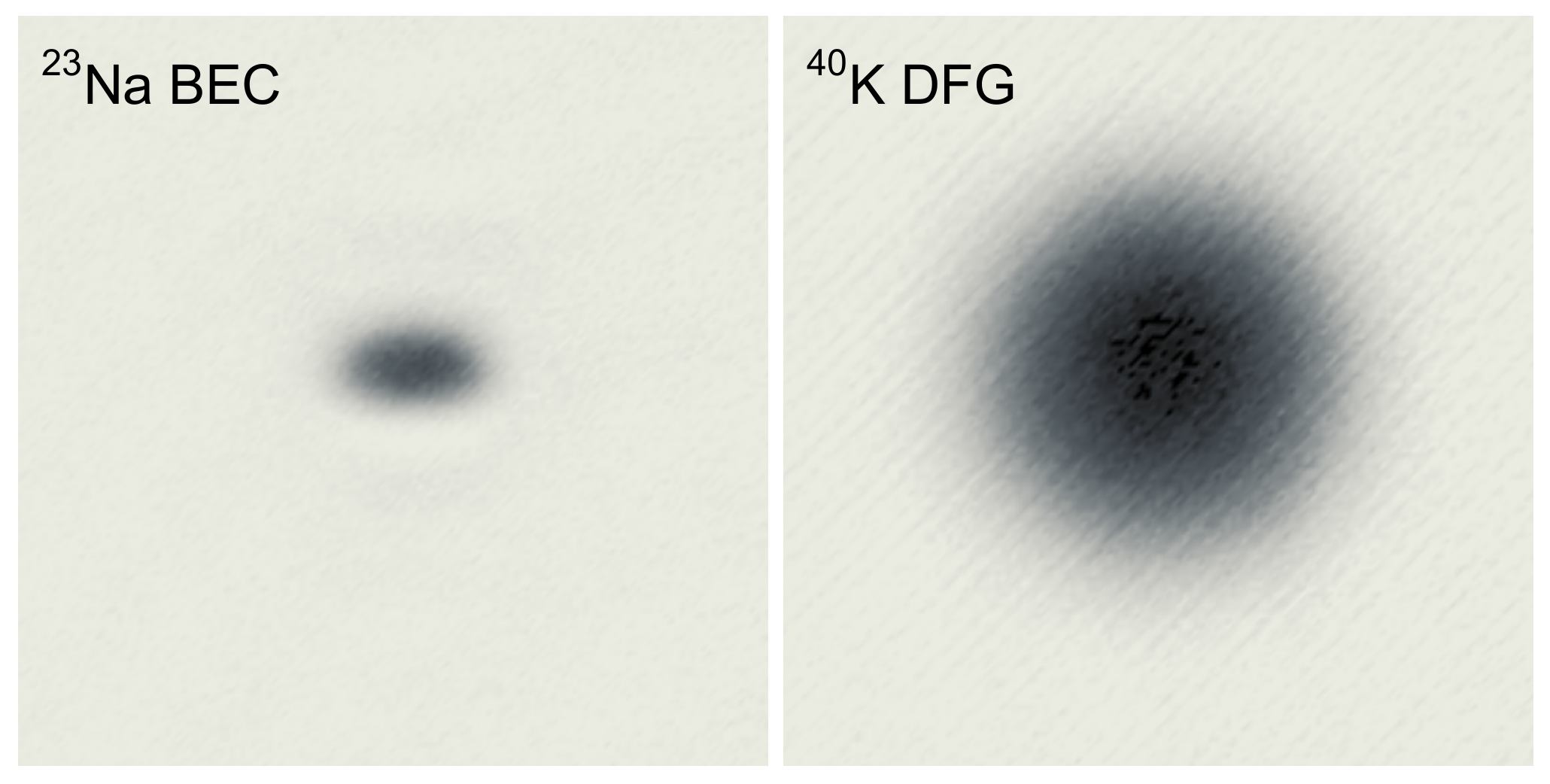}\caption{ Absorption images (average of 5 images) of a deeply degenerate Bose-Fermi mixture of $^{23}$Na and $^{40}$K atoms with a large number imbalance after a time-of-flight expansion of 13 ms. The mixture contains about $6.2\times10^4$ $^{23}$Na atoms with a Bose-Einstein condensate fraction of about 80\%  and about $3.9\times10^5 $ $^{40}$K atoms with $T/T_F=0.24$. }
\label{fig1}%
\end{figure}

Our experiment starts with the preparation of a deeply degenerate mixture of $^{23}$Na and $^{40}$K atoms. The experimental setup has been introduced in previous works \cite{Yang2022,Wang2022}. We first load $^{23}$Na and $^{40}$K atoms into a two-species dark-spot magneto-optical trap. The atoms are optically pumped to the $|f,m_{f}\rangle_{\rm{Na}}=|2,2\rangle$ state and the $|f,m_{f}\rangle_{\rm{K}}=|9/2,9/2\rangle$ state and are then loaded into a cloverleaf-type magnetic trap to perform evaporative cooling of $^{23}$Na atoms and the $^{40}$K atoms are sympathetically cooled. After that, the atomic mixtures are loaded into a horizontal dipole trap formed by two elliptical laser beams ($\lambda=1064$ nm) with beam waists of about $38\times152 $ $\mu$m crossed at an angle of about 5 degrees. The axial direction of the horizontal dipole trap is along the axial direction of the magnetic trap to ensure mode-matching with the cigar-shaped magnetic trap. In this horizontal dipole trap, $^{23}$Na atoms are transferred to the $|1,1\rangle$ state. A vertical dipole trap ($\lambda=1064$ nm)  with a beam waist of approximately $235$ $\mu$m is adiabatically ramped up to load the $^{23}$Na and $^{40}$K atoms into a large-volume dipole trap formed by the three laser beams. Optical evaporative cooling is performed in the large-volume three-beam optical dipole trap.  At the end of the optical evaporation, we typically create $6.2\times10^4$ $^{23}$Na atoms with a BEC fraction of about 80\% and about $3.9\times10^5 $ $^{40}$K atoms with a temperature of about 120 nK and $T/T_F\approx0.2-0.25$. The absorption images of the quantum degenerate atomic mixture are shown in Fig. \ref{fig1}. The trap frequencies for $^{40}$K atoms are $(\omega_x,\omega_y,\omega_z)=2 \pi\times (240,69,24)$ Hz.

A key improvement in the current work is that the interference between the two laser beams forming the horizontal dipole trap is largely suppressed. We find that the residual interference between these two laser beams causes additional heating. Although in our previous works a quantum degenerate atomic mixture can still be produced \cite{Wang2022}, a deeply degenerate atomic mixture with a large number imbalance was difficult. In the current work, the interference between the two beams is largely suppressed by carefully optimizing the polarizations and frequencies of the laser beams. This improvement and further optimization of the optical evaporative cooling allow us to create a deeply degenerate atomic mixture of $^{23}$Na and $^{40}$K atoms with a large number ratio ($N_F/N_B\approx 6$), which serves as an ideal starting point for the creation of quantum degenerate gas of molecules.

It is worth mentioning that in our work, we do not implement gray molasses on atoms. In Refs. \cite{DeMarco2019,Duda2021}, both the JILA and MPQ group implemented the gray molasses to improve the initial phase space density of the atoms. It was one of the key techniques for the preparation of a deeply degenerate mixture of $^{87}$Rb atoms and $^{40}$K atoms with a large number imbalance at the JILA group \cite{DeMarco2019}. Our work demonstrates that mode-matching loading from the magnetic trap into an optical dipole trap and evaporative cooling in a large-volume dipole trap can be used to prepare a deeply degenerate Bose-Fermi mixture with a large number ratio, even if the mixture has a large three-body loss rate.

After the preparation of the degenerate atomic mixture, we use the Feshbach resonance between the $|1,1\rangle$ and $|9/2,-7/2\rangle$ states at about 89.8 G to form $^{23}$Na$^{40}$K Feshbach molecules. To this end, we first transfer the $^{40}$K atoms to the $|9/2,-5/2\rangle$ state by a Landau-Zener passage at 32 G. We then ramp the magnetic field to 93 G and transfer the $^{40}$K atoms to the $|9/2,-7/2\rangle$ state by a $\pi$-pulse. The $^{23}$Na$^{40}$K Feshbach molecules are created by ramping the magnetic field from 93 G to 89.5 G at a rate of about 3 G/ms. We typically create about $4.3\times10^4$ $^{23}$Na$^{40}$K Feshbach molecules. The number of Feshbach molecules is determined by dissociating the Feshbach molecules into the $|1,1\rangle$+$|9/2,-9/2\rangle$ state, and the $^{40}$K atoms in the $|9/2,-9/2\rangle$ state are detected by absorption imaging. The efficiency of molecule creation for $^{23}$Na atoms is about 70\%. Note that this high association efficiency may be overestimated, because the thermal components of the $^{23}$Na atoms are difficult to discern for a nearly pure BEC and thus the number of $^{23}$Na atoms may be underestimated.

In the experiment at the MPQ group \cite{Duda2021}, high efficiency creation of $^{23}$Na$^{40}$K  Feshbach molecules can only be achieved by using a species-dependent optical dipole trap. However, we can achieve a high molecule creation efficiency in a conventional optical dipole trap, which is similar to the formation of $^{87}$Rb$^{40}$K molecules at the JILA group. This may be because the  large number ratio ($N_F/N_B\approx6$) in our work is close to that used in the JILA group  ($N_F/N_B\approx7$) \cite{DeMarco2019}. As pointed out in Refs. \cite{Cumby2013,DeMarco2019}, a large ratio of the number of fermionic atoms to bosonic atoms can help to mitigate the spatial density mismatch and suppress the three-body losses. In addition, the large volume dipole trap in our experiment also reduces the density of the atoms and thus can further suppress the three-body losses.

After the Feshbach molecules are created, we transfer the molecules to the rovibrational ground state by stimulated Raman adiabatic passage (STIRAP). We prepare the $^{23}$Na$^{40}$K molecules in the hyperfine level $|v,N,m_{i_{\rm{Na}}},m_{i_{\rm{K}}}\rangle=|0,0,-3/2,-3\rangle$ of the ground state, where $v$ and $N$ represent the vibrational and rotational quantum numbers, and $m_{i_{\rm{Na}}}$ and $m_{i_{\rm{K}}}$  denote the nuclear spin projections of $^{23}$Na and $^{40}$K, respectively. The $\sigma^{-}$  polarized pump laser and the $\sigma^{+}$ polarized Stokes laser propagate along the direction of the magnetic field. For a typical STIRAP efficiency of 70\%, we create about $3.0\times 10^4$ ground state molecules. To measure the degeneracy of the $^{23}$Na$^{40}$K molecules, we remove the remaining $^{23}$Na atoms and $^{40}$K atoms by resonant light pulses. After a short hold time, we transfer the ground-state molecules back to the Feshbach state for detection. The Feshbach molecules are detected by direct absorption imaging.

\begin{figure}[ptb]
\centering
\includegraphics[width=8cm]{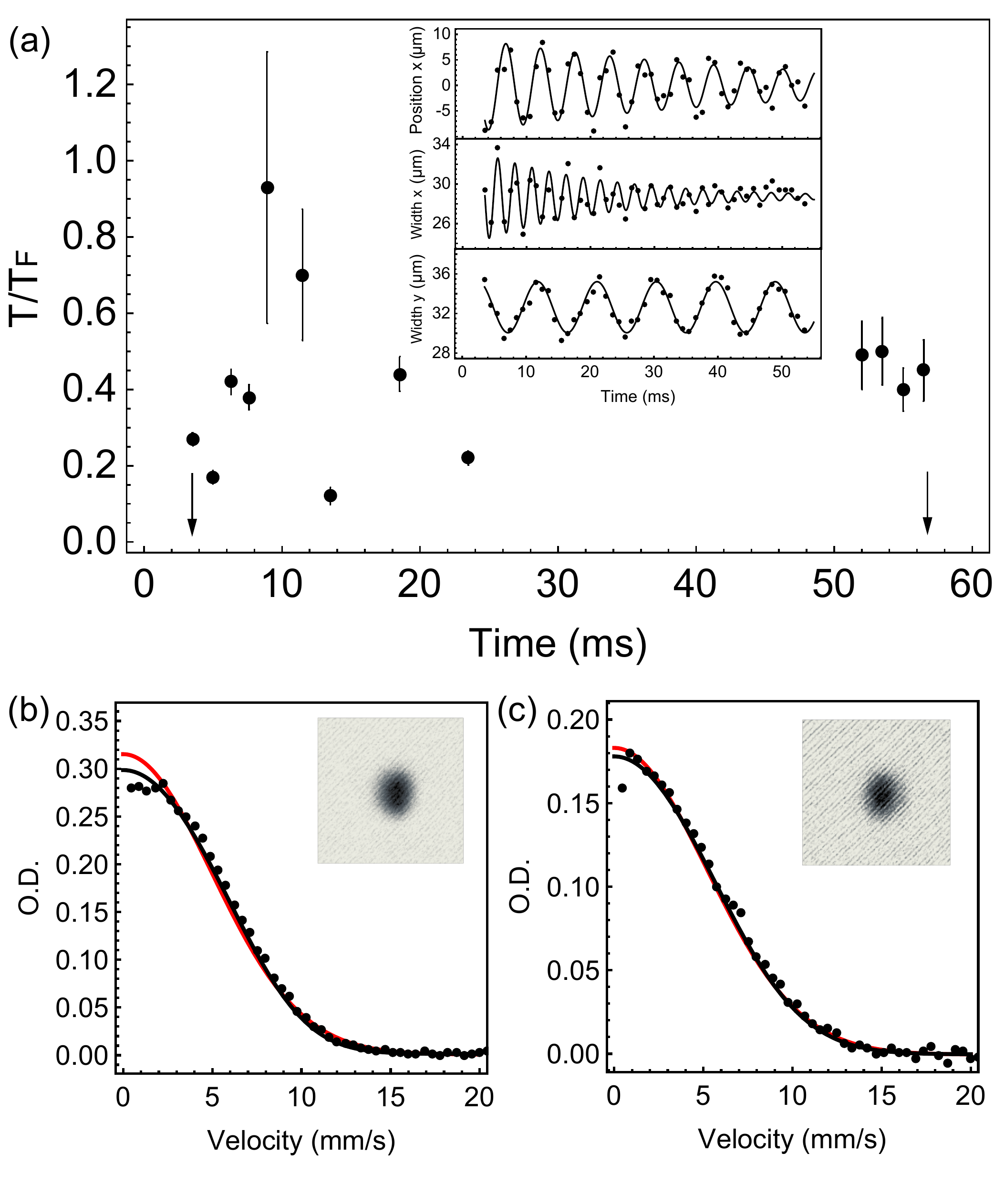}\caption{(a) The degeneracy parameter $T/T_F$ of the $^{23}$Na$^{40}$K molecules as a function of hold time. The 2D column density distribution (average of $10-12$ absorption images) is fit to the Fermi-Dirac distribution and $T/T_F$ is determined from the fitted fugacity.  The fluctuations of $T/T_F$ for a hold time less than about 25 ms are mainly affected by the breathing modes along the x direction. The collective modes of the molecules induced by STIRAP are shown in the inset, where the center-of-mass position of x direction and the widths of the molecular clouds along the x and y directions oscillate as a function of hold time. The $1/e$ time constants of the damping of the dipole oscillations and the breathing modes along the x direction are $46(12)$ ms and $19(4)$ ms, respectively. No notable damping is observed for the breathing mode along the y direction. The azimuthal averaged momentum distribution  for a hold time of $3.5$ ms and $56.5$ ms are shown in (b) and (c), respectively. The azimuthal averaging is applied after the fitting of the 2D distribution. The solid lines are the fit to the Fermi-Dirac distribution (black curve) and the Gaussian distribution (red curve). The inset shows the absorption images (average of (b)12 and (c)10 images) after a time-of-flight expansion of 9 ms. Error bars represent the standard error. }
\label{fig2}%
\end{figure}

The 2D column density distribution of the molecules is fit to the Fermi-Dirac distribution. The degeneracy parameter $T/T_F$ is obtained from the fitted fugacity and is shown in Fig. \ref{fig2}. We find that STIRAP excites the collective modes due to the misalignment and the differences of the trap frequencies. The dominant collective modes are the dipole modes along the x direction and the breathing modes in both the x and y directions. The degeneracy parameter $T/T_F$ changes significantly for a hold time of less than 25 ms, which is mainly affected by the breathing mode along the x direction. Due to the lack of a thermalization process, the collective oscillations can only be damped by the anharmonicity of the dipole trap. After a hold time of about 50 ms, the breathing mode along the x direction is damped while the breathing mode along the y direction remains. In this case, the amplitude that the degeneracy parameter fluctuates is small, and we obtain $T/T_F\approx0.4-0.5$. Since STIRAP conserves the shape of the molecule cloud, this value reflects the degeneracy of the Feshbach molecules. This value is smaller than the degeneracy parameter of $T/T_F\approx0.6$ calculated from the number of Feshbach molecules and the trap frequencies. This may be because the $^{23}$Na atoms are localized at the center of the $^{40}$K atoms, and the molecules are dominantly created from the low-entropy part of the deeply degenerate Fermi gas \cite{DeMarco2019}. Therefore, the Feshbach molecules inherit the low entropy of the $^{40}$K atoms. STIRAP introduces additional holes in the particle distribution due to the limited efficiency and thus reduces the degeneracy of the molecules. The effective degeneracy parameter of the ground-state molecules may be estimated from the modified peak occupancy \cite{Tobias2020}. For $T/T_F\approx0.4-0.5$ and a STIRAP efficiency of 70\%, the effective degeneracy parameter is $0.5-0.7$. However, a thermalization process is required for the ground-state molecules to reach thermal equilibrium.

After selectively removing the $^{23}$Na atoms, we obtain a quantum degenerate mixture of approximately $3\times10^4$ $^{23}$Na$^{40}$K molecules  and approximately $3\times10^5$ $^{40}$K atoms at a temperature of about 120 nK. The elastic collisions between $^{23}$Na$^{40}$K molecules and $^{40}$K atoms provide a thermalization mechanism for the molecules.
We measure the elastic scattering cross sections using the method in Ref. \cite{Su2022}, by monitoring the cross-species thermalization in a classical atom-molecule mixture. The measured elastic scattering cross section is $\sigma_{\rm{el}}=6.4(12)\times10^{-11}$ cm$^{2}$. The thermalization time  may be calculated using the formula $\tau_{\rm{th}}= (3/\xi)/\Gamma_{\rm{coll}}$, where $\Gamma_{\rm{coll}}=n_{\rm{ov}}\sigma_{\rm{el}} v_{\rm{rel}} $ is the elastic collision rate \cite{Mosk2001,Son2020,Tobias2020}. Here $n_{\rm{ov}}$ is the overlap density, $v_{\rm{rel}}$
is the relative mean velocity, and the parameter $\xi=4m_{\rm{NaK}}m_{\rm{K}}/(m_{\rm{NaK}}+m_{\rm{K}})^2$ represents the mass effect. For $T/T_F>0.2$, the Gaussian distribution is a good approximation for calculating $n_{\rm{ov}} v_{\rm{rel}}$ \cite{DeMarco2001}. Therefore, we have $n_{\rm{ov}}=(N_{\rm{NaK}}+N_{\rm{K}})[(\frac{2\pi k_{\rm{B}} T_{\rm{K}}}{m_{\rm{K}} \bar\omega_{\rm{K}}^2})(1+\frac{m_{\rm{K}}T_{\rm{NaK}}}{m_{\rm{NaK}} T_{\rm{K}} \gamma_t^2} )]^{-\frac{3}{2}}$ with $\gamma_t\approx0.77$ being the ratio between the trap frequencies
for $^{23}$Na$^{40}$K molecules and $^{40}$K atoms and $v_{\rm{rel}}=\sqrt{(8 k_{\rm{B}}/\pi)(T_{\rm{K}}/m_{\rm{K}}+T_{\rm{NaK}}/m_{\rm{NaK}})}$. For $T_{\rm{NaK}}=T_{\rm{K}}=120$ nK, $N_{\rm{K}}=3.0\times10^5$ and $N_{\rm{NaK}}=3.0\times10^4$, we obtain $\tau_{\rm{th}}\approx 7 $ ms.

These calculations do not consider the suppression of the elastic collisions due to the Pauli exclusion principle in the quantum degenerate regime, which is difficult to consider in the theoretical model. The effective elastic collision cross sections in the quantum degenerate regime may be obtained from the damping of the dipole oscillation in the mixture \cite{Tobias2020}. The damping rate of the dipole oscillations may be expressed in terms of the elastic collision rate as $1/\tau_{\rm{d}}-1/\tau_0=(2 m_{\rm{K}} N_{\rm{K}})/[3(m_{\rm{K}}+m_{\rm{NaK}})(N_{\rm{K}}+N_{\rm{NaK}})\Gamma_{\rm{coll}}]$, where $\tau_{\rm{d}}=17$ ms and $\tau_0=46$ ms are the measured $1/e$ time constants of damping of the dipole oscillations with and without atoms, respectively. We obtain an effective elastic scattering cross section $2.7(5)\times10^{-11}$ cm$^2/$. This value is smaller than that measured using the thermalization method in a classical mixture. We attribute this reduction to the suppression of the elastic collisions due to the Pauli exclusion principle \cite{DeMarco2001,DeMarcoth2001}. The thermalization time in the quantum degenerate regime may be increased by a similar factor, and thus we infer a thermalization time of about 17 ms, which is still much shorter than the lifetime of the molecules of about 85 ms in the mixtures.

\begin{figure}[ptb]
\centering
\includegraphics[width=8cm]{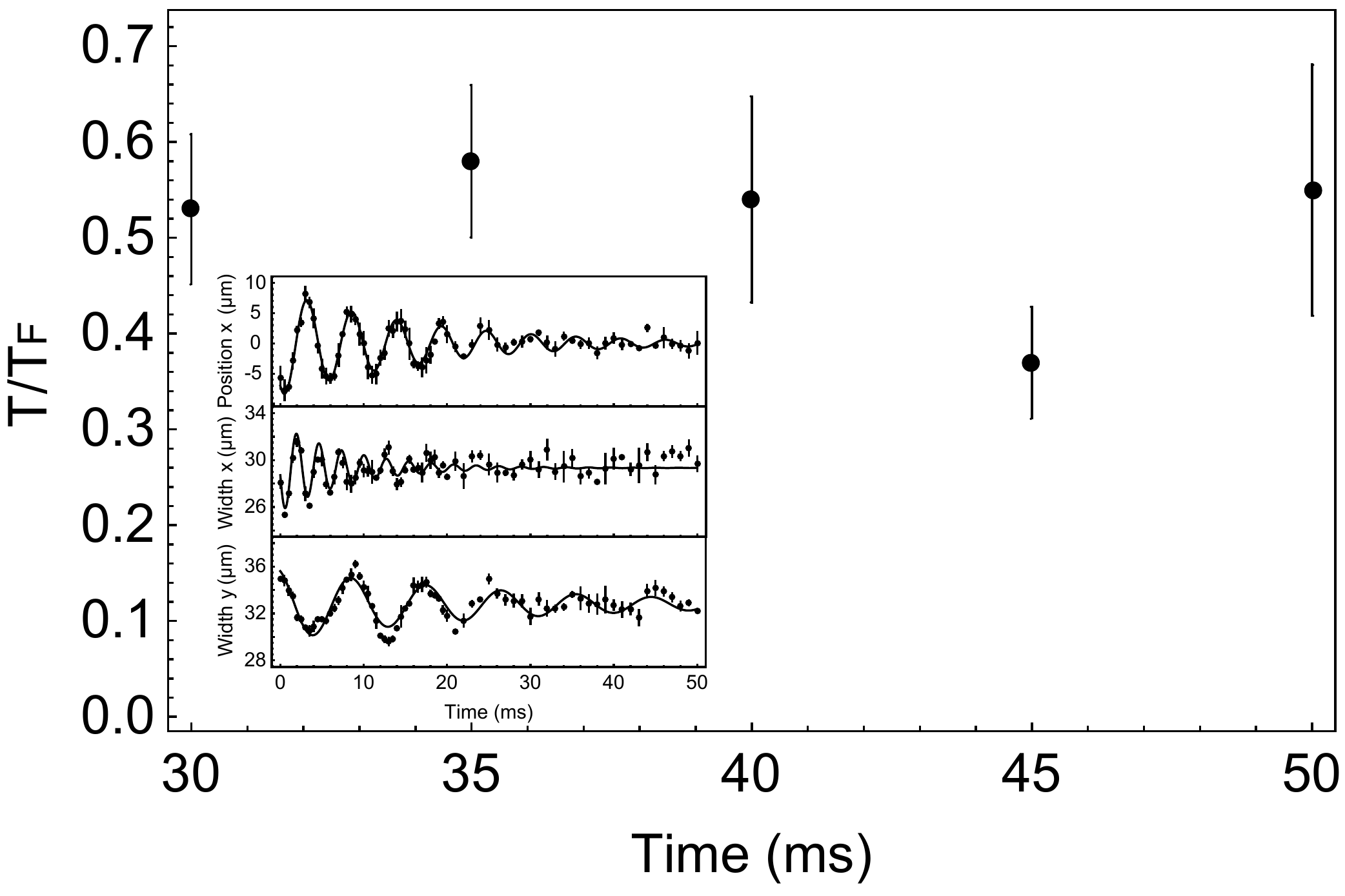}\caption{ The degeneracy parameter $T/T_F$ of the $^{23}$Na$^{40}$K molecules in the atom-molecule mixture as a function of hold time. The 2D column density distribution (average of $20-30$ images) is fit to the Fermi-Dirac distribution and the $T/T_F$ is determined from the fitted fugacity.  The damping of the collective oscillations of the molecules by the atom-molecule elastic collisions is shown in the inset. The $1/e$ time constants of the damping of the dipole oscillations and the breathing modes along the x and y directions are $17(2)$ ms, $8(2)$ ms, and $28(6)$ ms respectively. Error bars represent the standard error. }
\label{fig3}%
\end{figure}

In the mixtures, the breathing modes are also damped due to the elastic collisions. The damping time of the breathing mode is similar to that of the dipole oscillation. As shown in Fig. \ref{fig3}, we measure the degeneracy parameter of the molecules in the atom-molecule mixture for a hold time of $30-50$ ms, which is larger than the damping time of the collective oscillations and the estimated thermalization time. We find that the 2D column density distribution can still be fit to the Fermi-Dirac distribution with $T/T_F\approx0.5-0.6$. These degeneracy parameters are close to the effective degeneracy parameters estimated from the modified peak occupancy. However, they are smaller than the degeneracy parameters $T/T_F\approx0.8-1$ calculated from the trap frequencies and ground-state molecule numbers. A possible explanation is that increasing the degeneracy parameter $T/T_F$ requires absorbing energy from the atoms and creating more holes in the peak occupancy. However, absorbing energy from the more degenerate atoms is suppressed due to the Pauli exclusion principle \cite{DeMarco2001} and thus a complete thermalization process may need a longer time than the thermalization time that the simple thermalization model predicts. On the timescale of a few tens of milliseconds, the degeneracy of the molecules can be maintained, which is favorable for the study of the quantum degenerate atom-molecule mixtures.

In summary, we have prepared a quantum degenerate mixture of $^{23}$Na$^{40}$K molecules and $^{40}$K atoms. The method of preparing the degenerate molecules in our work may be useful to other atomic mixtures that also have large three-body loss rates. The collective oscillations induced by the STIRAP lasers can be suppressed by using 1D optical lattices \cite{DeMarco2019} and optimizing the alignment. Further increasing the degeneracy may be achieved by selectively removing the high-energy $^{40}$K atoms, and using the atom-molecule elastic collisions to perform evaporative cooling. There are many Feshbach resonances between $^{23}$Na$^{40}$K molecules and $^{40}$K atoms. Therefore, the preparation of the quantum degenerate mixture with tunable interactions paves the way toward studying strongly interacting atom-molecule mixtures, quantum simulation of fermionic angulon \cite{Schmidt2015,Schmidt2016}, and preparation of triatomic molecular gases \cite{Trenkwalder2011,Koehler2006,Hermsmeier2021}. The mixture may also be used to study the BCS pairing of atoms and molecules, similar to the ultracold mixture of mass-imbalanced fermionic atoms \cite{Ravensbergen2020,Green2020}.


\begin{acknowledgments}
This work was supported by the National  Key R\&D Program of China (under Grant No.
2018YFA0306502), the National Natural Science Foundation of China (under Grant Nos. 11521063, 11904355),  the
Chinese Academy of Sciences, the Anhui Initiative in Quantum Information Technologies, the Shanghai Municipal Science and Technology Major Project (Grant No.2019SHZDZX01), the Shanghai Rising-Star Program (Grant No. 20QA1410000), the Innovation Program for Quantum Science and Technology (Grant No. 2021ZD0302101).
\end{acknowledgments}


\begin{thebibliography}{47}
\expandafter\ifx\csname natexlab\endcsname\relax\def\natexlab#1{#1}\fi
\expandafter\ifx\csname bibnamefont\endcsname\relax
  \def\bibnamefont#1{#1}\fi
\expandafter\ifx\csname bibfnamefont\endcsname\relax
  \def\bibfnamefont#1{#1}\fi
\expandafter\ifx\csname citenamefont\endcsname\relax
  \def\citenamefont#1{#1}\fi
\expandafter\ifx\csname url\endcsname\relax
  \def\url#1{\texttt{#1}}\fi
\expandafter\ifx\csname urlprefix\endcsname\relax\def\urlprefix{URL }\fi
\providecommand{\bibinfo}[2]{#2}
\providecommand{\eprint}[2][]{\url{#2}}

\bibitem[{\citenamefont{Krems}(2008)}]{Krems2008}
\bibinfo{author}{\bibfnamefont{R.~V.} \bibnamefont{Krems}},
  \bibinfo{journal}{Phys. Chem. Chem. Phys.} \textbf{\bibinfo{volume}{10}},
  \bibinfo{pages}{4079} (\bibinfo{year}{2008}).

\bibitem[{\citenamefont{Qu\'{e}m\'{e}ner and Julienne}(2012)}]{Quemener2012}
\bibinfo{author}{\bibfnamefont{G.}~\bibnamefont{Qu\'{e}m\'{e}ner}}
  \bibnamefont{and} \bibinfo{author}{\bibfnamefont{P.~S.}
  \bibnamefont{Julienne}}, \bibinfo{journal}{Chem. Rev.}
  \textbf{\bibinfo{volume}{112}}, \bibinfo{pages}{4949} (\bibinfo{year}{2012}).

\bibitem[{\citenamefont{Carr et~al.}(2009)\citenamefont{Carr, DeMille, Krems,
  and Ye}}]{Carr2009}
\bibinfo{author}{\bibfnamefont{L.~D.} \bibnamefont{Carr}},
  \bibinfo{author}{\bibfnamefont{D.}~\bibnamefont{DeMille}},
  \bibinfo{author}{\bibfnamefont{R.~V.} \bibnamefont{Krems}}, \bibnamefont{and}
  \bibinfo{author}{\bibfnamefont{J.}~\bibnamefont{Ye}}, \bibinfo{journal}{New
  J. Phys.} \textbf{\bibinfo{volume}{11}}, \bibinfo{pages}{055049}
  (\bibinfo{year}{2009}).

\bibitem[{\citenamefont{Yang et~al.}(2019)\citenamefont{Yang, Zhang, Liu, Liu,
  Nan, Zhao, and Pan}}]{Yang2019}
\bibinfo{author}{\bibfnamefont{H.}~\bibnamefont{Yang}},
  \bibinfo{author}{\bibfnamefont{D.-C.} \bibnamefont{Zhang}},
  \bibinfo{author}{\bibfnamefont{L.}~\bibnamefont{Liu}},
  \bibinfo{author}{\bibfnamefont{Y.-X.} \bibnamefont{Liu}},
  \bibinfo{author}{\bibfnamefont{J.}~\bibnamefont{Nan}},
  \bibinfo{author}{\bibfnamefont{B.}~\bibnamefont{Zhao}}, \bibnamefont{and}
  \bibinfo{author}{\bibfnamefont{J.-W.} \bibnamefont{Pan}},
  \bibinfo{journal}{Science} \textbf{\bibinfo{volume}{363}},
  \bibinfo{pages}{261} (\bibinfo{year}{2019}).

\bibitem[{\citenamefont{Yang et~al.}(2022)\citenamefont{Yang, Wang, Su, Cao,
  Zhang, Rui, Zhao, Bai, and Pan}}]{Yang2022}
\bibinfo{author}{\bibfnamefont{H.}~\bibnamefont{Yang}},
  \bibinfo{author}{\bibfnamefont{X.-Y.} \bibnamefont{Wang}},
  \bibinfo{author}{\bibfnamefont{Z.}~\bibnamefont{Su}},
  \bibinfo{author}{\bibfnamefont{J.}~\bibnamefont{Cao}},
  \bibinfo{author}{\bibfnamefont{D.-C.} \bibnamefont{Zhang}},
  \bibinfo{author}{\bibfnamefont{J.}~\bibnamefont{Rui}},
  \bibinfo{author}{\bibfnamefont{B.}~\bibnamefont{Zhao}},
  \bibinfo{author}{\bibfnamefont{C.-L.} \bibnamefont{Bai}}, \bibnamefont{and}
  \bibinfo{author}{\bibfnamefont{J.-W.} \bibnamefont{Pan}},
  \bibinfo{journal}{Nature} \textbf{\bibinfo{volume}{602}},
  \bibinfo{pages}{229} (\bibinfo{year}{2022}).

\bibitem[{\citenamefont{Wang et~al.}(2021)\citenamefont{Wang, Frye, Su, Cao,
  Liu, Zhang, Yang, Hutson, Zhao, Bai et~al.}}]{Wang2021}
\bibinfo{author}{\bibfnamefont{X.-Y.} \bibnamefont{Wang}},
  \bibinfo{author}{\bibfnamefont{M.~D.} \bibnamefont{Frye}},
  \bibinfo{author}{\bibfnamefont{Z.}~\bibnamefont{Su}},
  \bibinfo{author}{\bibfnamefont{J.}~\bibnamefont{Cao}},
  \bibinfo{author}{\bibfnamefont{L.}~\bibnamefont{Liu}},
  \bibinfo{author}{\bibfnamefont{D.-C.} \bibnamefont{Zhang}},
  \bibinfo{author}{\bibfnamefont{H.}~\bibnamefont{Yang}},
  \bibinfo{author}{\bibfnamefont{J.~M.} \bibnamefont{Hutson}},
  \bibinfo{author}{\bibfnamefont{B.}~\bibnamefont{Zhao}},
  \bibinfo{author}{\bibfnamefont{C.-L.} \bibnamefont{Bai}},
  \bibnamefont{et~al.}, \bibinfo{journal}{New J. Phys.}
  \textbf{\bibinfo{volume}{23}}, \bibinfo{pages}{115010}
  (\bibinfo{year}{2021}).

\bibitem[{\citenamefont{Su et~al.}(2022)\citenamefont{Su, Yang, Cao, Wang, Rui,
  Zhao, and Pan}}]{Su2022}
\bibinfo{author}{\bibfnamefont{Z.}~\bibnamefont{Su}},
  \bibinfo{author}{\bibfnamefont{H.}~\bibnamefont{Yang}},
  \bibinfo{author}{\bibfnamefont{J.}~\bibnamefont{Cao}},
  \bibinfo{author}{\bibfnamefont{X.-Y.} \bibnamefont{Wang}},
  \bibinfo{author}{\bibfnamefont{J.}~\bibnamefont{Rui}},
  \bibinfo{author}{\bibfnamefont{B.}~\bibnamefont{Zhao}}, \bibnamefont{and}
  \bibinfo{author}{\bibfnamefont{J.-W.} \bibnamefont{Pan}},
  \bibinfo{journal}{Phys. Rev. Lett.} \textbf{\bibinfo{volume}{129}},
  \bibinfo{pages}{033401} (\bibinfo{year}{2022}).

\bibitem[{\citenamefont{Son et~al.}(2022)\citenamefont{Son, Park, Lu, Jamison,
  Karman, and Ketterle}}]{Son2022}
\bibinfo{author}{\bibfnamefont{H.}~\bibnamefont{Son}},
  \bibinfo{author}{\bibfnamefont{J.~J.} \bibnamefont{Park}},
  \bibinfo{author}{\bibfnamefont{Y.-K.} \bibnamefont{Lu}},
  \bibinfo{author}{\bibfnamefont{A.~O.} \bibnamefont{Jamison}},
  \bibinfo{author}{\bibfnamefont{T.}~\bibnamefont{Karman}}, \bibnamefont{and}
  \bibinfo{author}{\bibfnamefont{W.}~\bibnamefont{Ketterle}},
  \bibinfo{journal}{Science} \textbf{\bibinfo{volume}{375}},
  \bibinfo{pages}{1006} (\bibinfo{year}{2022}).

\bibitem[{\citenamefont{Chin et~al.}(2010)\citenamefont{Chin, Grimm, Julienne,
  and Tiesinga}}]{chin2010}
\bibinfo{author}{\bibfnamefont{C.}~\bibnamefont{Chin}},
  \bibinfo{author}{\bibfnamefont{R.}~\bibnamefont{Grimm}},
  \bibinfo{author}{\bibfnamefont{P.}~\bibnamefont{Julienne}}, \bibnamefont{and}
  \bibinfo{author}{\bibfnamefont{E.}~\bibnamefont{Tiesinga}},
  \bibinfo{journal}{Rev. Mod. Phys.} \textbf{\bibinfo{volume}{82}},
  \bibinfo{pages}{1225} (\bibinfo{year}{2010}).

\bibitem[{\citenamefont{K\"ohler et~al.}(2006)\citenamefont{K\"ohler, G\'oral,
  and Julienne}}]{Koehler2006}
\bibinfo{author}{\bibfnamefont{T.}~\bibnamefont{K\"ohler}},
  \bibinfo{author}{\bibfnamefont{K.}~\bibnamefont{G\'oral}}, \bibnamefont{and}
  \bibinfo{author}{\bibfnamefont{P.~S.} \bibnamefont{Julienne}},
  \bibinfo{journal}{Rev. Mod. Phys.} \textbf{\bibinfo{volume}{78}},
  \bibinfo{pages}{1311} (\bibinfo{year}{2006}).

\bibitem[{\citenamefont{Hermsmeier et~al.}(2021)\citenamefont{Hermsmeier,
  K\l{}os, Kotochigova, and Tscherbul}}]{Hermsmeier2021}
\bibinfo{author}{\bibfnamefont{R.}~\bibnamefont{Hermsmeier}},
  \bibinfo{author}{\bibfnamefont{J.}~\bibnamefont{K\l{}os}},
  \bibinfo{author}{\bibfnamefont{S.}~\bibnamefont{Kotochigova}},
  \bibnamefont{and} \bibinfo{author}{\bibfnamefont{T.~V.}
  \bibnamefont{Tscherbul}}, \bibinfo{journal}{Phys. Rev. Lett.}
  \textbf{\bibinfo{volume}{127}}, \bibinfo{pages}{103402}
  (\bibinfo{year}{2021}).

\bibitem[{\citenamefont{Kraemer et~al.}(2006)\citenamefont{Kraemer, Mark,
  Waldburger, Danzl, Chin, Engeser, Lange, Pilch, Jaakkola, N{\"a}gerl
  et~al.}}]{Kraemer2006}
\bibinfo{author}{\bibfnamefont{T.}~\bibnamefont{Kraemer}},
  \bibinfo{author}{\bibfnamefont{M.}~\bibnamefont{Mark}},
  \bibinfo{author}{\bibfnamefont{P.}~\bibnamefont{Waldburger}},
  \bibinfo{author}{\bibfnamefont{J.~G.} \bibnamefont{Danzl}},
  \bibinfo{author}{\bibfnamefont{C.}~\bibnamefont{Chin}},
  \bibinfo{author}{\bibfnamefont{B.}~\bibnamefont{Engeser}},
  \bibinfo{author}{\bibfnamefont{A.~D.} \bibnamefont{Lange}},
  \bibinfo{author}{\bibfnamefont{K.}~\bibnamefont{Pilch}},
  \bibinfo{author}{\bibfnamefont{A.}~\bibnamefont{Jaakkola}},
  \bibinfo{author}{\bibfnamefont{H.-C.} \bibnamefont{N{\"a}gerl}},
  \bibnamefont{et~al.}, \bibinfo{journal}{Nature}
  \textbf{\bibinfo{volume}{440}}, \bibinfo{pages}{315} (\bibinfo{year}{2006}).

\bibitem[{\citenamefont{Schmidt and Lemeshko}(2015)}]{Schmidt2015}
\bibinfo{author}{\bibfnamefont{R.}~\bibnamefont{Schmidt}} \bibnamefont{and}
  \bibinfo{author}{\bibfnamefont{M.}~\bibnamefont{Lemeshko}},
  \bibinfo{journal}{Phys. Rev. Lett.} \textbf{\bibinfo{volume}{114}},
  \bibinfo{pages}{203001} (\bibinfo{year}{2015}).

\bibitem[{\citenamefont{Schmidt and Lemeshko}(2016)}]{Schmidt2016}
\bibinfo{author}{\bibfnamefont{R.}~\bibnamefont{Schmidt}} \bibnamefont{and}
  \bibinfo{author}{\bibfnamefont{M.}~\bibnamefont{Lemeshko}},
  \bibinfo{journal}{Phys. Rev. X} \textbf{\bibinfo{volume}{6}},
  \bibinfo{pages}{011012} (\bibinfo{year}{2016}).

\bibitem[{\citenamefont{Ni et~al.}(2008)\citenamefont{Ni, Ospelkaus,
  de~Miranda, Pe'er, Neyenhuis, Zirbel, Kotochigova, Julienne, Jin, and
  Ye}}]{Ni2008}
\bibinfo{author}{\bibfnamefont{K.-K.} \bibnamefont{Ni}},
  \bibinfo{author}{\bibfnamefont{S.}~\bibnamefont{Ospelkaus}},
  \bibinfo{author}{\bibfnamefont{M.~H.~G.} \bibnamefont{de~Miranda}},
  \bibinfo{author}{\bibfnamefont{A.}~\bibnamefont{Pe'er}},
  \bibinfo{author}{\bibfnamefont{B.}~\bibnamefont{Neyenhuis}},
  \bibinfo{author}{\bibfnamefont{J.~J.} \bibnamefont{Zirbel}},
  \bibinfo{author}{\bibfnamefont{S.}~\bibnamefont{Kotochigova}},
  \bibinfo{author}{\bibfnamefont{P.~S.} \bibnamefont{Julienne}},
  \bibinfo{author}{\bibfnamefont{D.~S.} \bibnamefont{Jin}}, \bibnamefont{and}
  \bibinfo{author}{\bibfnamefont{J.}~\bibnamefont{Ye}},
  \bibinfo{journal}{Science} \textbf{\bibinfo{volume}{322}},
  \bibinfo{pages}{231} (\bibinfo{year}{2008}).

\bibitem[{\citenamefont{Molony et~al.}(2014)\citenamefont{Molony, Gregory, Ji,
  Lu, K\"oppinger, Le~Sueur, Blackley, Hutson, and Cornish}}]{Molony2014}
\bibinfo{author}{\bibfnamefont{P.~K.} \bibnamefont{Molony}},
  \bibinfo{author}{\bibfnamefont{P.~D.} \bibnamefont{Gregory}},
  \bibinfo{author}{\bibfnamefont{Z.}~\bibnamefont{Ji}},
  \bibinfo{author}{\bibfnamefont{B.}~\bibnamefont{Lu}},
  \bibinfo{author}{\bibfnamefont{M.~P.} \bibnamefont{K\"oppinger}},
  \bibinfo{author}{\bibfnamefont{C.~R.} \bibnamefont{Le~Sueur}},
  \bibinfo{author}{\bibfnamefont{C.~L.} \bibnamefont{Blackley}},
  \bibinfo{author}{\bibfnamefont{J.~M.} \bibnamefont{Hutson}},
  \bibnamefont{and} \bibinfo{author}{\bibfnamefont{S.~L.}
  \bibnamefont{Cornish}}, \bibinfo{journal}{Phys. Rev. Lett.}
  \textbf{\bibinfo{volume}{113}}, \bibinfo{pages}{255301}
  (\bibinfo{year}{2014}).

\bibitem[{\citenamefont{Takekoshi et~al.}(2014)\citenamefont{Takekoshi,
  Reichs\"ollner, Schindewolf, Hutson, Le~Sueur, Dulieu, Ferlaino, Grimm, and
  N\"agerl}}]{Takekoshi2014}
\bibinfo{author}{\bibfnamefont{T.}~\bibnamefont{Takekoshi}},
  \bibinfo{author}{\bibfnamefont{L.}~\bibnamefont{Reichs\"ollner}},
  \bibinfo{author}{\bibfnamefont{A.}~\bibnamefont{Schindewolf}},
  \bibinfo{author}{\bibfnamefont{J.~M.} \bibnamefont{Hutson}},
  \bibinfo{author}{\bibfnamefont{C.~R.} \bibnamefont{Le~Sueur}},
  \bibinfo{author}{\bibfnamefont{O.}~\bibnamefont{Dulieu}},
  \bibinfo{author}{\bibfnamefont{F.}~\bibnamefont{Ferlaino}},
  \bibinfo{author}{\bibfnamefont{R.}~\bibnamefont{Grimm}}, \bibnamefont{and}
  \bibinfo{author}{\bibfnamefont{H.-C.} \bibnamefont{N\"agerl}},
  \bibinfo{journal}{Phys. Rev. Lett.} \textbf{\bibinfo{volume}{113}},
  \bibinfo{pages}{205301} (\bibinfo{year}{2014}).

\bibitem[{\citenamefont{Park et~al.}(2015)\citenamefont{Park, Will, and
  Zwierlein}}]{Park2015}
\bibinfo{author}{\bibfnamefont{J.~W.} \bibnamefont{Park}},
  \bibinfo{author}{\bibfnamefont{S.~A.} \bibnamefont{Will}}, \bibnamefont{and}
  \bibinfo{author}{\bibfnamefont{M.~W.} \bibnamefont{Zwierlein}},
  \bibinfo{journal}{Phys. Rev. Lett.} \textbf{\bibinfo{volume}{114}},
  \bibinfo{pages}{205302} (\bibinfo{year}{2015}).

\bibitem[{\citenamefont{Guo et~al.}(2016)\citenamefont{Guo, Zhu, Lu, Ye, Wang,
  Vexiau, Bouloufa-Maafa, Qu\'em\'ener, Dulieu, and Wang}}]{Guo2016}
\bibinfo{author}{\bibfnamefont{M.}~\bibnamefont{Guo}},
  \bibinfo{author}{\bibfnamefont{B.}~\bibnamefont{Zhu}},
  \bibinfo{author}{\bibfnamefont{B.}~\bibnamefont{Lu}},
  \bibinfo{author}{\bibfnamefont{X.}~\bibnamefont{Ye}},
  \bibinfo{author}{\bibfnamefont{F.}~\bibnamefont{Wang}},
  \bibinfo{author}{\bibfnamefont{R.}~\bibnamefont{Vexiau}},
  \bibinfo{author}{\bibfnamefont{N.}~\bibnamefont{Bouloufa-Maafa}},
  \bibinfo{author}{\bibfnamefont{G.}~\bibnamefont{Qu\'em\'ener}},
  \bibinfo{author}{\bibfnamefont{O.}~\bibnamefont{Dulieu}}, \bibnamefont{and}
  \bibinfo{author}{\bibfnamefont{D.}~\bibnamefont{Wang}},
  \bibinfo{journal}{Phys. Rev. Lett.} \textbf{\bibinfo{volume}{116}},
  \bibinfo{pages}{205303} (\bibinfo{year}{2016}).

\bibitem[{\citenamefont{Rvachov et~al.}(2017)\citenamefont{Rvachov, Son,
  Sommer, Ebadi, Park, Zwierlein, Ketterle, and Jamison}}]{Rvachov2017}
\bibinfo{author}{\bibfnamefont{T.~M.} \bibnamefont{Rvachov}},
  \bibinfo{author}{\bibfnamefont{H.}~\bibnamefont{Son}},
  \bibinfo{author}{\bibfnamefont{A.~T.} \bibnamefont{Sommer}},
  \bibinfo{author}{\bibfnamefont{S.}~\bibnamefont{Ebadi}},
  \bibinfo{author}{\bibfnamefont{J.~J.} \bibnamefont{Park}},
  \bibinfo{author}{\bibfnamefont{M.~W.} \bibnamefont{Zwierlein}},
  \bibinfo{author}{\bibfnamefont{W.}~\bibnamefont{Ketterle}}, \bibnamefont{and}
  \bibinfo{author}{\bibfnamefont{A.~O.} \bibnamefont{Jamison}},
  \bibinfo{journal}{Phys. Rev. Lett.} \textbf{\bibinfo{volume}{119}},
  \bibinfo{pages}{143001} (\bibinfo{year}{2017}).

\bibitem[{\citenamefont{See\ss{}elberg
  et~al.}(2018)\citenamefont{See\ss{}elberg, Buchheim, Lu, Schneider, Luo,
  Tiemann, Bloch, and Gohle}}]{seesselberg2018}
\bibinfo{author}{\bibfnamefont{F.}~\bibnamefont{See\ss{}elberg}},
  \bibinfo{author}{\bibfnamefont{N.}~\bibnamefont{Buchheim}},
  \bibinfo{author}{\bibfnamefont{Z.-K.} \bibnamefont{Lu}},
  \bibinfo{author}{\bibfnamefont{T.}~\bibnamefont{Schneider}},
  \bibinfo{author}{\bibfnamefont{X.-Y.} \bibnamefont{Luo}},
  \bibinfo{author}{\bibfnamefont{E.}~\bibnamefont{Tiemann}},
  \bibinfo{author}{\bibfnamefont{I.}~\bibnamefont{Bloch}}, \bibnamefont{and}
  \bibinfo{author}{\bibfnamefont{C.}~\bibnamefont{Gohle}},
  \bibinfo{journal}{Phys. Rev. A} \textbf{\bibinfo{volume}{97}},
  \bibinfo{pages}{013405} (\bibinfo{year}{2018}).

\bibitem[{\citenamefont{Liu et~al.}(2019)\citenamefont{Liu, Zhang, Yang, Liu,
  Nan, Rui, Zhao, and Pan}}]{LiuL2019}
\bibinfo{author}{\bibfnamefont{L.}~\bibnamefont{Liu}},
  \bibinfo{author}{\bibfnamefont{D.-C.} \bibnamefont{Zhang}},
  \bibinfo{author}{\bibfnamefont{H.}~\bibnamefont{Yang}},
  \bibinfo{author}{\bibfnamefont{Y.-X.} \bibnamefont{Liu}},
  \bibinfo{author}{\bibfnamefont{J.}~\bibnamefont{Nan}},
  \bibinfo{author}{\bibfnamefont{J.}~\bibnamefont{Rui}},
  \bibinfo{author}{\bibfnamefont{B.}~\bibnamefont{Zhao}}, \bibnamefont{and}
  \bibinfo{author}{\bibfnamefont{J.-W.} \bibnamefont{Pan}},
  \bibinfo{journal}{Phys. Rev. Lett.} \textbf{\bibinfo{volume}{122}},
  \bibinfo{pages}{253201} (\bibinfo{year}{2019}).

\bibitem[{\citenamefont{Voges et~al.}(2020)\citenamefont{Voges, Gersema,
  Meyer~zum Alten~Borgloh, Schulze, Hartmann, Zenesini, and
  Ospelkaus}}]{Voges2020}
\bibinfo{author}{\bibfnamefont{K.~K.} \bibnamefont{Voges}},
  \bibinfo{author}{\bibfnamefont{P.}~\bibnamefont{Gersema}},
  \bibinfo{author}{\bibfnamefont{M.}~\bibnamefont{Meyer~zum Alten~Borgloh}},
  \bibinfo{author}{\bibfnamefont{T.~A.} \bibnamefont{Schulze}},
  \bibinfo{author}{\bibfnamefont{T.}~\bibnamefont{Hartmann}},
  \bibinfo{author}{\bibfnamefont{A.}~\bibnamefont{Zenesini}}, \bibnamefont{and}
  \bibinfo{author}{\bibfnamefont{S.}~\bibnamefont{Ospelkaus}},
  \bibinfo{journal}{Phys. Rev. Lett.} \textbf{\bibinfo{volume}{125}},
  \bibinfo{pages}{083401} (\bibinfo{year}{2020}).

\bibitem[{\citenamefont{Barry et~al.}(2014)\citenamefont{Barry, McCarron,
  Norrgard, Steinecker, and DeMille}}]{Barry2014}
\bibinfo{author}{\bibfnamefont{J.~F.} \bibnamefont{Barry}},
  \bibinfo{author}{\bibfnamefont{D.~J.} \bibnamefont{McCarron}},
  \bibinfo{author}{\bibfnamefont{E.~B.} \bibnamefont{Norrgard}},
  \bibinfo{author}{\bibfnamefont{M.~H.} \bibnamefont{Steinecker}},
  \bibnamefont{and} \bibinfo{author}{\bibfnamefont{D.}~\bibnamefont{DeMille}},
  \bibinfo{journal}{Nature} \textbf{\bibinfo{volume}{512}},
  \bibinfo{pages}{286} (\bibinfo{year}{2014}).

\bibitem[{\citenamefont{Cheuk et~al.}(2018)\citenamefont{Cheuk, Anderegg,
  Augenbraun, Bao, Burchesky, Ketterle, and Doyle}}]{Cheuk2018}
\bibinfo{author}{\bibfnamefont{L.~W.} \bibnamefont{Cheuk}},
  \bibinfo{author}{\bibfnamefont{L.}~\bibnamefont{Anderegg}},
  \bibinfo{author}{\bibfnamefont{B.~L.} \bibnamefont{Augenbraun}},
  \bibinfo{author}{\bibfnamefont{Y.}~\bibnamefont{Bao}},
  \bibinfo{author}{\bibfnamefont{S.}~\bibnamefont{Burchesky}},
  \bibinfo{author}{\bibfnamefont{W.}~\bibnamefont{Ketterle}}, \bibnamefont{and}
  \bibinfo{author}{\bibfnamefont{J.~M.} \bibnamefont{Doyle}},
  \bibinfo{journal}{Phys. Rev. Lett.} \textbf{\bibinfo{volume}{121}},
  \bibinfo{pages}{083201} (\bibinfo{year}{2018}).

\bibitem[{\citenamefont{Caldwell et~al.}(2019)\citenamefont{Caldwell, Devlin,
  Williams, Fitch, Hinds, Sauer, and Tarbutt}}]{Caldwell2019}
\bibinfo{author}{\bibfnamefont{L.}~\bibnamefont{Caldwell}},
  \bibinfo{author}{\bibfnamefont{J.~A.} \bibnamefont{Devlin}},
  \bibinfo{author}{\bibfnamefont{H.~J.} \bibnamefont{Williams}},
  \bibinfo{author}{\bibfnamefont{N.~J.} \bibnamefont{Fitch}},
  \bibinfo{author}{\bibfnamefont{E.~A.} \bibnamefont{Hinds}},
  \bibinfo{author}{\bibfnamefont{B.~E.} \bibnamefont{Sauer}}, \bibnamefont{and}
  \bibinfo{author}{\bibfnamefont{M.~R.} \bibnamefont{Tarbutt}},
  \bibinfo{journal}{Phys. Rev. Lett.} \textbf{\bibinfo{volume}{123}},
  \bibinfo{pages}{033202} (\bibinfo{year}{2019}).

\bibitem[{\citenamefont{Ding et~al.}(2020)\citenamefont{Ding, Wu, Finneran,
  Burau, and Ye}}]{Ding2020}
\bibinfo{author}{\bibfnamefont{S.}~\bibnamefont{Ding}},
  \bibinfo{author}{\bibfnamefont{Y.}~\bibnamefont{Wu}},
  \bibinfo{author}{\bibfnamefont{I.~A.} \bibnamefont{Finneran}},
  \bibinfo{author}{\bibfnamefont{J.~J.} \bibnamefont{Burau}}, \bibnamefont{and}
  \bibinfo{author}{\bibfnamefont{J.}~\bibnamefont{Ye}}, \bibinfo{journal}{Phys.
  Rev. X} \textbf{\bibinfo{volume}{10}}, \bibinfo{pages}{021049}
  (\bibinfo{year}{2020}).

\bibitem[{\citenamefont{DeMarco et~al.}(2019)\citenamefont{DeMarco, Valtolina,
  Matsuda, Tobias, Covey, and Ye}}]{DeMarco2019}
\bibinfo{author}{\bibfnamefont{L.}~\bibnamefont{DeMarco}},
  \bibinfo{author}{\bibfnamefont{G.}~\bibnamefont{Valtolina}},
  \bibinfo{author}{\bibfnamefont{K.}~\bibnamefont{Matsuda}},
  \bibinfo{author}{\bibfnamefont{W.~G.} \bibnamefont{Tobias}},
  \bibinfo{author}{\bibfnamefont{J.~P.} \bibnamefont{Covey}}, \bibnamefont{and}
  \bibinfo{author}{\bibfnamefont{J.}~\bibnamefont{Ye}},
  \bibinfo{journal}{Science} \textbf{\bibinfo{volume}{363}},
  \bibinfo{pages}{853} (\bibinfo{year}{2019}).

\bibitem[{\citenamefont{Valtolina et~al.}(2020)\citenamefont{Valtolina,
  Matsuda, Tobias, Li, Marco, and Ye}}]{Giacomo2020}
\bibinfo{author}{\bibfnamefont{G.}~\bibnamefont{Valtolina}},
  \bibinfo{author}{\bibfnamefont{K.}~\bibnamefont{Matsuda}},
  \bibinfo{author}{\bibfnamefont{W.~G.} \bibnamefont{Tobias}},
  \bibinfo{author}{\bibfnamefont{J.-R.} \bibnamefont{Li}},
  \bibinfo{author}{\bibfnamefont{L.~D.} \bibnamefont{Marco}}, \bibnamefont{and}
  \bibinfo{author}{\bibfnamefont{J.}~\bibnamefont{Ye}},
  \bibinfo{journal}{Nature} \textbf{\bibinfo{volume}{588}},
  \bibinfo{pages}{239} (\bibinfo{year}{2020}).

\bibitem[{\citenamefont{Duda et~al.}(2021)\citenamefont{Duda, Chen,
  Schindewolf, Bause, von Milczewski, Schmidt, Bloch, and Luo}}]{Duda2021}
\bibinfo{author}{\bibfnamefont{M.}~\bibnamefont{Duda}},
  \bibinfo{author}{\bibfnamefont{X.-Y.} \bibnamefont{Chen}},
  \bibinfo{author}{\bibfnamefont{A.}~\bibnamefont{Schindewolf}},
  \bibinfo{author}{\bibfnamefont{R.}~\bibnamefont{Bause}},
  \bibinfo{author}{\bibfnamefont{J.}~\bibnamefont{von Milczewski}},
  \bibinfo{author}{\bibfnamefont{R.}~\bibnamefont{Schmidt}},
  \bibinfo{author}{\bibfnamefont{I.}~\bibnamefont{Bloch}}, \bibnamefont{and}
  \bibinfo{author}{\bibfnamefont{X.-Y.} \bibnamefont{Luo}},
  \bibinfo{journal}{arXiv: 2111.04301v2}  (\bibinfo{year}{2021}).

\bibitem[{\citenamefont{Schindewolf et~al.}(2022)\citenamefont{Schindewolf,
  Bause, Chen, Duda, Karman, Bloch, and Luo}}]{Schindewolf2022}
\bibinfo{author}{\bibfnamefont{A.}~\bibnamefont{Schindewolf}},
  \bibinfo{author}{\bibfnamefont{R.}~\bibnamefont{Bause}},
  \bibinfo{author}{\bibfnamefont{X.-Y.} \bibnamefont{Chen}},
  \bibinfo{author}{\bibfnamefont{M.}~\bibnamefont{Duda}},
  \bibinfo{author}{\bibfnamefont{T.}~\bibnamefont{Karman}},
  \bibinfo{author}{\bibfnamefont{I.}~\bibnamefont{Bloch}}, \bibnamefont{and}
  \bibinfo{author}{\bibfnamefont{X.-Y.} \bibnamefont{Luo}},
  \bibinfo{journal}{Nature} \textbf{\bibinfo{volume}{607}},
  \bibinfo{pages}{677} (\bibinfo{year}{2022}).

\bibitem[{\citenamefont{Christianen et~al.}(2019)\citenamefont{Christianen,
  Zwierlein, Groenenboom, and Karman}}]{Christianen2019}
\bibinfo{author}{\bibfnamefont{A.}~\bibnamefont{Christianen}},
  \bibinfo{author}{\bibfnamefont{M.~W.} \bibnamefont{Zwierlein}},
  \bibinfo{author}{\bibfnamefont{G.~C.} \bibnamefont{Groenenboom}},
  \bibnamefont{and} \bibinfo{author}{\bibfnamefont{T.}~\bibnamefont{Karman}},
  \bibinfo{journal}{Phys. Rev. Lett.} \textbf{\bibinfo{volume}{123}},
  \bibinfo{pages}{123402} (\bibinfo{year}{2019}).

\bibitem[{\citenamefont{Liu et~al.}(2020)\citenamefont{Liu, Hu, Nichols,
  Grimes, Karman, Guo, and Ni}}]{Liu2020}
\bibinfo{author}{\bibfnamefont{Y.}~\bibnamefont{Liu}},
  \bibinfo{author}{\bibfnamefont{M.-G.} \bibnamefont{Hu}},
  \bibinfo{author}{\bibfnamefont{M.~A.} \bibnamefont{Nichols}},
  \bibinfo{author}{\bibfnamefont{D.~D.} \bibnamefont{Grimes}},
  \bibinfo{author}{\bibfnamefont{T.}~\bibnamefont{Karman}},
  \bibinfo{author}{\bibfnamefont{H.}~\bibnamefont{Guo}}, \bibnamefont{and}
  \bibinfo{author}{\bibfnamefont{K.-K.} \bibnamefont{Ni}},
  \bibinfo{journal}{Nat. Phys.} \textbf{\bibinfo{volume}{16}},
  \bibinfo{pages}{1132} (\bibinfo{year}{2020}).

\bibitem[{\citenamefont{Gregory et~al.}(2020)\citenamefont{Gregory, Blackmore,
  Bromley, and Cornish}}]{Gregory2020}
\bibinfo{author}{\bibfnamefont{P.~D.} \bibnamefont{Gregory}},
  \bibinfo{author}{\bibfnamefont{J.~A.} \bibnamefont{Blackmore}},
  \bibinfo{author}{\bibfnamefont{S.~L.} \bibnamefont{Bromley}},
  \bibnamefont{and} \bibinfo{author}{\bibfnamefont{S.~L.}
  \bibnamefont{Cornish}}, \bibinfo{journal}{Phys. Rev. Lett.}
  \textbf{\bibinfo{volume}{124}}, \bibinfo{pages}{163402}
  (\bibinfo{year}{2020}).

\bibitem[{\citenamefont{Nichols et~al.}(2022)\citenamefont{Nichols, Liu, Zhu,
  Hu, Liu, and Ni}}]{Nichols2022}
\bibinfo{author}{\bibfnamefont{M.~A.} \bibnamefont{Nichols}},
  \bibinfo{author}{\bibfnamefont{Y.-X.} \bibnamefont{Liu}},
  \bibinfo{author}{\bibfnamefont{L.}~\bibnamefont{Zhu}},
  \bibinfo{author}{\bibfnamefont{M.-G.} \bibnamefont{Hu}},
  \bibinfo{author}{\bibfnamefont{Y.}~\bibnamefont{Liu}}, \bibnamefont{and}
  \bibinfo{author}{\bibfnamefont{K.-K.} \bibnamefont{Ni}},
  \bibinfo{journal}{Phys. Rev. X} \textbf{\bibinfo{volume}{12}},
  \bibinfo{pages}{011049} (\bibinfo{year}{2022}).

\bibitem[{\citenamefont{Mayle et~al.}(2012)\citenamefont{Mayle, Ruzic, and
  Bohn}}]{Mayle2012}
\bibinfo{author}{\bibfnamefont{M.}~\bibnamefont{Mayle}},
  \bibinfo{author}{\bibfnamefont{B.~P.} \bibnamefont{Ruzic}}, \bibnamefont{and}
  \bibinfo{author}{\bibfnamefont{J.~L.} \bibnamefont{Bohn}},
  \bibinfo{journal}{Phys. Rev. A} \textbf{\bibinfo{volume}{85}},
  \bibinfo{pages}{062712} (\bibinfo{year}{2012}).

\bibitem[{\citenamefont{Frye and Hutson}(2021)}]{Matthew2021}
\bibinfo{author}{\bibfnamefont{M.~D.} \bibnamefont{Frye}} \bibnamefont{and}
  \bibinfo{author}{\bibfnamefont{J.~M.} \bibnamefont{Hutson}},
  \bibinfo{journal}{New J. Phys.} \textbf{\bibinfo{volume}{23}},
  \bibinfo{pages}{125008} (\bibinfo{year}{2021}).

\bibitem[{\citenamefont{Wang et~al.}(2022)\citenamefont{Wang, Su, Cao, Yang,
  Zhao, Bai, and Pan}}]{Wang2022}
\bibinfo{author}{\bibfnamefont{X.-Y.} \bibnamefont{Wang}},
  \bibinfo{author}{\bibfnamefont{Z.}~\bibnamefont{Su}},
  \bibinfo{author}{\bibfnamefont{J.}~\bibnamefont{Cao}},
  \bibinfo{author}{\bibfnamefont{H.}~\bibnamefont{Yang}},
  \bibinfo{author}{\bibfnamefont{B.}~\bibnamefont{Zhao}},
  \bibinfo{author}{\bibfnamefont{C.-L.} \bibnamefont{Bai}}, \bibnamefont{and}
  \bibinfo{author}{\bibfnamefont{J.-W.} \bibnamefont{Pan}},
  \bibinfo{journal}{Sci. China-Phys. Mech. Astron.}
  \textbf{\bibinfo{volume}{65}}, \bibinfo{pages}{223011}
  (\bibinfo{year}{2022}).

\bibitem[{\citenamefont{Cumby et~al.}(2013)\citenamefont{Cumby, Shewmon, Hu,
  Perreault, and Jin}}]{Cumby2013}
\bibinfo{author}{\bibfnamefont{T.~D.} \bibnamefont{Cumby}},
  \bibinfo{author}{\bibfnamefont{R.~A.} \bibnamefont{Shewmon}},
  \bibinfo{author}{\bibfnamefont{M.-G.} \bibnamefont{Hu}},
  \bibinfo{author}{\bibfnamefont{J.~D.} \bibnamefont{Perreault}},
  \bibnamefont{and} \bibinfo{author}{\bibfnamefont{D.~S.} \bibnamefont{Jin}},
  \bibinfo{journal}{Phys. Rev. A} \textbf{\bibinfo{volume}{87}},
  \bibinfo{pages}{012703} (\bibinfo{year}{2013}).

\bibitem[{\citenamefont{Tobias et~al.}(2020)\citenamefont{Tobias, Matsuda,
  Valtolina, De~Marco, Li, and Ye}}]{Tobias2020}
\bibinfo{author}{\bibfnamefont{W.~G.} \bibnamefont{Tobias}},
  \bibinfo{author}{\bibfnamefont{K.}~\bibnamefont{Matsuda}},
  \bibinfo{author}{\bibfnamefont{G.}~\bibnamefont{Valtolina}},
  \bibinfo{author}{\bibfnamefont{L.}~\bibnamefont{De~Marco}},
  \bibinfo{author}{\bibfnamefont{J.-R.} \bibnamefont{Li}}, \bibnamefont{and}
  \bibinfo{author}{\bibfnamefont{J.}~\bibnamefont{Ye}}, \bibinfo{journal}{Phys.
  Rev. Lett.} \textbf{\bibinfo{volume}{124}}, \bibinfo{pages}{033401}
  (\bibinfo{year}{2020}).

\bibitem[{\citenamefont{Mosk et~al.}(2001)\citenamefont{Mosk, Kraft, Mudrich,
  Singer, Wohlleben, Grimm, and Weidem\"{u}ller}}]{Mosk2001}
\bibinfo{author}{\bibfnamefont{A.}~\bibnamefont{Mosk}},
  \bibinfo{author}{\bibfnamefont{S.}~\bibnamefont{Kraft}},
  \bibinfo{author}{\bibfnamefont{M.}~\bibnamefont{Mudrich}},
  \bibinfo{author}{\bibfnamefont{K.}~\bibnamefont{Singer}},
  \bibinfo{author}{\bibfnamefont{W.}~\bibnamefont{Wohlleben}},
  \bibinfo{author}{\bibfnamefont{R.}~\bibnamefont{Grimm}}, \bibnamefont{and}
  \bibinfo{author}{\bibfnamefont{M.}~\bibnamefont{Weidem\"{u}ller}},
  \bibinfo{journal}{Applied Physics B} \textbf{\bibinfo{volume}{73}},
  \bibinfo{pages}{791} (\bibinfo{year}{2001}).

\bibitem[{\citenamefont{Son et~al.}(2020)\citenamefont{Son, Park, Ketterle, and
  Jamison}}]{Son2020}
\bibinfo{author}{\bibfnamefont{H.}~\bibnamefont{Son}},
  \bibinfo{author}{\bibfnamefont{J.~J.} \bibnamefont{Park}},
  \bibinfo{author}{\bibfnamefont{W.}~\bibnamefont{Ketterle}}, \bibnamefont{and}
  \bibinfo{author}{\bibfnamefont{A.~O.} \bibnamefont{Jamison}},
  \bibinfo{journal}{Nature} \textbf{\bibinfo{volume}{580}},
  \bibinfo{pages}{197} (\bibinfo{year}{2020}).

\bibitem[{\citenamefont{DeMarco et~al.}(2001)\citenamefont{DeMarco, Papp, and
  Jin}}]{DeMarco2001}
\bibinfo{author}{\bibfnamefont{B.}~\bibnamefont{DeMarco}},
  \bibinfo{author}{\bibfnamefont{S.~B.} \bibnamefont{Papp}}, \bibnamefont{and}
  \bibinfo{author}{\bibfnamefont{D.~S.} \bibnamefont{Jin}},
  \bibinfo{journal}{Phys. Rev. Lett.} \textbf{\bibinfo{volume}{86}},
  \bibinfo{pages}{5409} (\bibinfo{year}{2001}).

\bibitem[{\citenamefont{DeMarco}(2001)}]{DeMarcoth2001}
\bibinfo{author}{\bibfnamefont{B.}~\bibnamefont{DeMarco}}, Ph.D. thesis,
  \bibinfo{school}{University of Colorado} (\bibinfo{year}{2001}).

\bibitem[{\citenamefont{Trenkwalder et~al.}(2011)\citenamefont{Trenkwalder,
  Kohstall, Zaccanti, Naik, Sidorov, Schreck, and Grimm}}]{Trenkwalder2011}
\bibinfo{author}{\bibfnamefont{A.}~\bibnamefont{Trenkwalder}},
  \bibinfo{author}{\bibfnamefont{C.}~\bibnamefont{Kohstall}},
  \bibinfo{author}{\bibfnamefont{M.}~\bibnamefont{Zaccanti}},
  \bibinfo{author}{\bibfnamefont{D.}~\bibnamefont{Naik}},
  \bibinfo{author}{\bibfnamefont{A.~I.} \bibnamefont{Sidorov}},
  \bibinfo{author}{\bibfnamefont{F.}~\bibnamefont{Schreck}}, \bibnamefont{and}
  \bibinfo{author}{\bibfnamefont{R.}~\bibnamefont{Grimm}},
  \bibinfo{journal}{Phys. Rev. Lett.} \textbf{\bibinfo{volume}{106}},
  \bibinfo{pages}{115304} (\bibinfo{year}{2011}).

\bibitem[{\citenamefont{Ravensbergen et~al.}(2020)\citenamefont{Ravensbergen,
  Soave, Corre, Kreyer, Huang, Kirilov, and Grimm}}]{Ravensbergen2020}
\bibinfo{author}{\bibfnamefont{C.}~\bibnamefont{Ravensbergen}},
  \bibinfo{author}{\bibfnamefont{E.}~\bibnamefont{Soave}},
  \bibinfo{author}{\bibfnamefont{V.}~\bibnamefont{Corre}},
  \bibinfo{author}{\bibfnamefont{M.}~\bibnamefont{Kreyer}},
  \bibinfo{author}{\bibfnamefont{B.}~\bibnamefont{Huang}},
  \bibinfo{author}{\bibfnamefont{E.}~\bibnamefont{Kirilov}}, \bibnamefont{and}
  \bibinfo{author}{\bibfnamefont{R.}~\bibnamefont{Grimm}},
  \bibinfo{journal}{Phys. Rev. Lett.} \textbf{\bibinfo{volume}{124}},
  \bibinfo{pages}{203402} (\bibinfo{year}{2020}).

\bibitem[{\citenamefont{Green et~al.}(2020)\citenamefont{Green, Li, See~Toh,
  Tang, McCormick, Li, Tiesinga, Kotochigova, and Gupta}}]{Green2020}
\bibinfo{author}{\bibfnamefont{A.}~\bibnamefont{Green}},
  \bibinfo{author}{\bibfnamefont{H.}~\bibnamefont{Li}},
  \bibinfo{author}{\bibfnamefont{J.~H.} \bibnamefont{See~Toh}},
  \bibinfo{author}{\bibfnamefont{X.}~\bibnamefont{Tang}},
  \bibinfo{author}{\bibfnamefont{K.~C.} \bibnamefont{McCormick}},
  \bibinfo{author}{\bibfnamefont{M.}~\bibnamefont{Li}},
  \bibinfo{author}{\bibfnamefont{E.}~\bibnamefont{Tiesinga}},
  \bibinfo{author}{\bibfnamefont{S.}~\bibnamefont{Kotochigova}},
  \bibnamefont{and} \bibinfo{author}{\bibfnamefont{S.}~\bibnamefont{Gupta}},
  \bibinfo{journal}{Phys. Rev. X} \textbf{\bibinfo{volume}{10}},
  \bibinfo{pages}{031037} (\bibinfo{year}{2020}).

\end{thebibliography}
\end{document}